\documentclass[aps,prl,twocolumn,groupedaddress]{revtex4-1}
\usepackage{amsmath}
\usepackage{amssymb}
\usepackage{graphicx}
\usepackage{dcolumn}
\usepackage{bm}

\begin{document}

\title{Fourier-component engineering to control light diffraction beyond subwavelength limit}
%\title{Mechanisms to control light diffraction beyond the subwavelength limit}

\author{Sun-Goo Lee}
\email{sungooleee@gmail.com}
\author{Seong-Han Kim}
\author{Chul-Sik Kee}
\email{cskee@gist.ac.kr}
\affiliation{Integrated Optics Laboratory, Advanced Photonics Research Institute, Gwangju Institute of Science and Technology, Gwangju 61005, South Korea}
\date{\today}

\begin{abstract}
In conventional diffraction theory, a subwavelength period is considered a prerequisite to achieve interesting resonance-assisted physical phenomena, such as bound states in the continuum and diverse zero-order spectral responses with $100\%$ diffraction efficiency. Here, we present modified diffraction equations that provide mechanisms to control light diffraction beyond the subwavelength limit. We show that resonant diffraction phenomena are governed by the superposition of scattering processes, owing to higher Fourier harmonic components. By appropriately engineering the Fourier harmonic components in the grating parameters, unwanted diffraction orders can be suppressed. Moreover, bound states in the continuum and highly efficient zero-order spectral responses can be achieved beyond the subwavelength limit. The concept of engineering Fourier harmonic components in periodic modulations provides new mechanisms to overcome the diffraction limit.
\end{abstract}

%\pacs{78.67-n, 42.70.Qs}

\maketitle

The analysis of light diffracted by gratings has a history of more than 200 years. Since the pioneering studies by Young and Fraunhofer \cite{TYoung1802,JFraunhofer1823}, frequency-selective functionalities have been realized by utilizing higher diffraction orders when the period of the grating ($\Lambda$) is larger than the wavelength of incident light ($\lambda$). The directions of the diffracted light can be predicted from the grating equation \cite{NBonod2016}. However, the discovery of Wood's anomalies in 1902 \cite{RWWood1902} prompted the study and development of subwavelength ($\Lambda < \lambda$) resonance-assisted gratings, which are typically composed of a waveguide and periodic diffracting elements and are of great scientific interest. Incident light is captured by resonances through Bloch waveguide modes and reemitted to predesigned spectral responses with a $100\%$ diffraction efficiency \cite{YHKo2018}. Resonance properties are governed by the eigenfrequencies of the Bloch modes in the gratings \cite{Magnusson2009}. Owing to recent advances in numerical capabilities and fabrication technologies, different types of resonance-assisted gratings have been presented for a variety of applications \cite{FBruckner2010,CJChang-Hasnain2012,RMagnusson2014,JWYoon2015,MNiraula2015,GQuaranta2018,Hemmati2019,XYin2020}.

Recent studies suggest that the eigenfrequencies of the Bloch modes in resonance-assisted gratings with slab geometry are governed by the interplay between higher Fourier harmonic components in the lattice parameters \cite{SGLee2019-1,SGLee2019-2,SGLee2020-2}. Below the diffraction limit, the radiative properties of Bloch modes can be controlled by engineering the Fourier harmonic components \cite{SGLee2021-1,SGLee2021-2}. In conventional diffraction theory, the subwavelength period is an essential condition for achieving important resonance-assisted physical phenomena, such as bound states in the continuum (BICs) and diverse zero-order spectral responses with $100\%$ diffraction efficiency \cite{BZhen2014,XGao2016,LNi2016,SDKrasikov2018,ENBulgakov2018,SGLee2020-1,AIOvcharenko2020,DNMaksimov2020}. Beyond the subwavelength limit $\Lambda > \lambda$, higher diffraction orders provide additional radiation channels and make it difficult to achieve $100\%$ diffraction efficiency in the zero-order direction.

In this study, we present new grating equations, namely resonance-assisted grating equations, that provide powerful mechanisms to control light diffraction beyond the subwavelength limit. We also show that the resonant diffraction phenomenon is the superimposed effect of scattering processes, owing to higher Fourier harmonic components in periodic modulations. We demonstrate that both BICs and highly efficient zero-order spectral responses can be achieved, even beyond the subwavelength limit, by suppressing unwanted diffraction orders via the engineering of Fourier harmonic components.

\begin{figure}[t]
\centering \includegraphics[width=8.5cm]{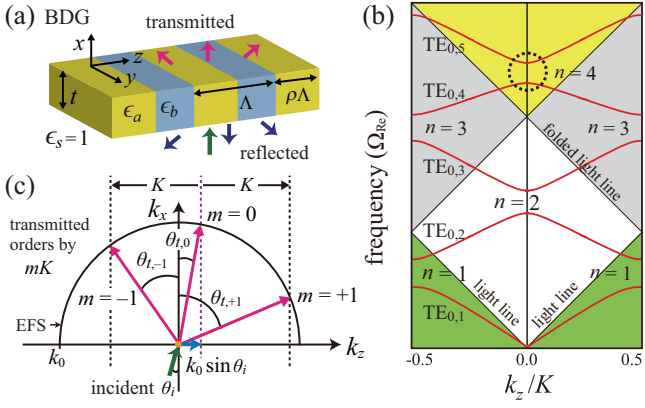}
\caption {\label{fig1} Light diffraction by a resonance-assisted grating. (a) Schematic of a conventional BDG that supports multiple diffraction orders. (b) Dispersion curves for fundamental $\mathrm{TE}_{0}$ mode. $K=2 \pi /\Lambda$ is the magnitude of the grating vector. (c) Diffraction orders obtained by the conventional grating equation near the fourth stop band (black dotted circle).}
\end{figure}

Figure~\ref{fig1}(a) illustrates the simplest representative resonant-assisted grating, i.e., a binary dielectric grating (BDG). It comprises high- ($\epsilon_a$) and low- ($\epsilon_b$) dielectric constant materials and is enclosed by a surrounding medium ($\epsilon_s =1$). The thickness of the grating is $t$, and the width of the high-dielectric-constant section is $\rho \Lambda$. The grating layer supports transverse electric ($\mathrm{TE}_{q}$) guided modes because its average dielectric constant $\epsilon_{\rm{av}}=\epsilon_{b}+\rho (\epsilon_{a}- \epsilon_{b})$ is larger than $\epsilon_s$. With spatial dielectric constant modulation, Bloch mode dispersion relations are conventionally represented in the irreducible Brillouin zone. Figure~\ref{fig1}(b) schematically illustrates dispersion curves for the fundamental $\mathrm{TE}_{0}$ mode. Between guided bands $\mathrm{TE}_{0,n}$ and $\mathrm{TE}_{0,n+1}$, the $n$th photonic band gap opens via the Bragg reflection process. Guided modes in the green region are not associated with the diffraction effect because they are protected by total internal reflection. However, those in the white, grey, and yellow regions generate diffracted light because they are described by the complex frequency $\Omega = \Omega_{\mathrm{Re}} +i\Omega_{\mathrm{Im}}$. There are numerous studies that only consider zero-order diffraction effects in the white subwavelength region. For the first time, to our knowledge, this study considers the resonant diffractions by the fundamental $\mathrm{TE}_{0}$ modes in the grey and yellow regions beyond the subwavelength limit. We first investigate the resonant diffractions at the fourth and sixth stop bands, which are associated with the normal incidence of light, and we then extend the analysis to the third stop band.

When a light beam of frequency $\Omega_{\mathrm{Re}} =k_0/K$ is incident on a BDG with an angle of $\theta_i$, waveguide modes in the vicinity of the fourth stop band in the yellow region generate diffracted beams with orders $m=-1$, $0$, and $+1$. Typically, the directions of the $m$th-order diffracted light can be precisely determined using the conventional grating equation:
\begin{equation}\label{phase-matching}
k_0 \sin \theta_{r,m} = k_0 \sin \theta_{t,m} = \Delta k_z  + mK,
\end{equation} 	 	
where $\theta_{r,m} $ and $\theta_{t,m}$ represent the angles of the reflected and transmitted beams, respectively, and $\Delta k_z = k_0 \sin \theta_{i}$. As illustrated in Fig.~\ref{fig1}(c), diffraction orders can be visualized by utilizing the equifrequency surface (EFS) of the surrounding medium \cite{MNotomi2000}. Because $\theta_{r,m} = \theta_{t,m}$ for a BDG surrounded by a single transparent medium, only transmitted orders are illustrated. In this conventional picture, diffraction orders are explained well through the additional momentum $mK$, owing to the grating with a period of $\Lambda$. However, the diffraction processes in Eq.~(\ref{phase-matching}) do not describe the roles of individual higher Fourier harmonic components with a period of $\Lambda_{n \geq 1}=\Lambda/n$. To include the contributions of individual higher Fourier harmonic components, we expand the periodic dielectric function as a Fourier cosine function series $\epsilon(z)=\sum_{0}^{\infty} \epsilon_{n}\cos(nKz)$, where the Fourier coefficients are given by $ \epsilon_{0}=\epsilon_{\rm{av}}$ and $ \epsilon_{n\geq1}=(2\Delta \epsilon / n\pi) \sin (n\pi \rho)$. Additionally, we consider the scattering effects owing to higher Fourier harmonics. In periodic photonic structures, the electric field is generally expressed in Bloch form. For the guided modes near the fourth stop band that open at the third-order $\Gamma$ point ($k_z=2~K$ in the extended Brillouin zone), the field distribution can be expressed as
\begin{equation}\label{electric-field}
\begin{split}
E_{y}(x,z)&=A\exp[i( \Delta k_z+ 2K )z]\varphi(x) \\
               &+ B\exp[i( \Delta k_z- 2K )z]\varphi(x) + E_{\rm{rad}},
 \end{split}
\end{equation} 	 	
where $A$ and $B$ are slowly-varying envelopes of the two counter-propagating waves, $\varphi(x)$ characterizes the mode profile of the unmodulated waveguide, and $\Delta k_z = k_0 \sin \theta_{i}$ is the deviation from $k_z=0$ \cite{Kazarinov1985,Rosenblatt1997,YDing2007}. We consider that the two counter-propagating waves, $P(x,z)=A\exp[i( \Delta k_z+ 2K )z]\varphi(x)$ and $N(x,z)=B\exp[i( \Delta k_z- 2K )z]\varphi(x)$, produce the diffracting wave $E_{\rm{rad}}$ via the superposition of scattering processes, owing to higher Fourier harmonics $\epsilon_{n\geq1}\cos(nKz)$.

\begin{figure}[]
\centering \includegraphics[width=8.0cm]{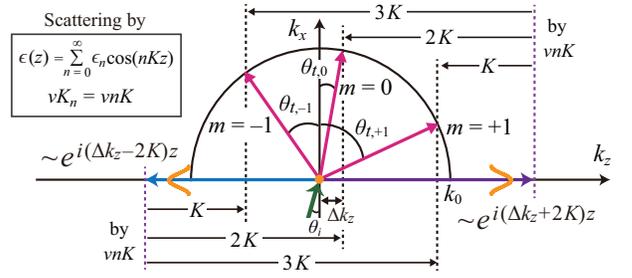}
\caption {\label{fig2} Conceptual illustration of the resonant diffraction by Eq.~(\ref{modified-phase-matching}) in the vicinity of the fourth stop band beyond the subwavelength limit.}
\end{figure}

\begin{figure*}[]
\centering \includegraphics[width=17.5 cm]{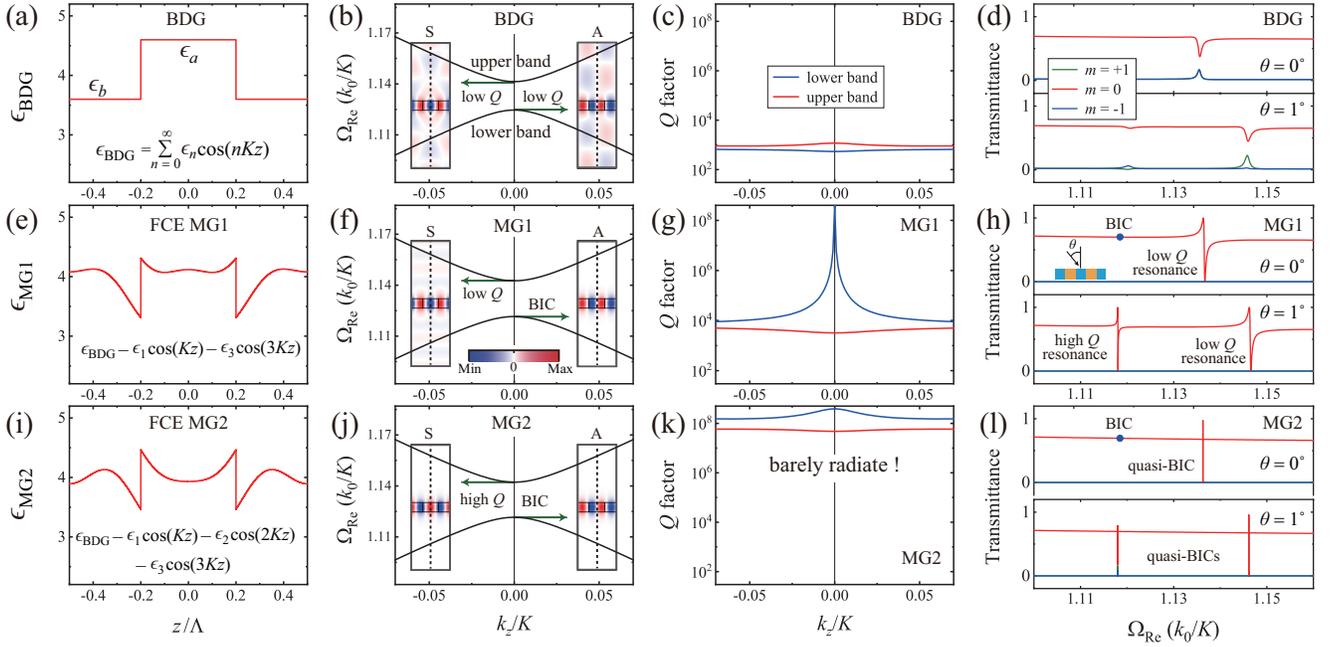}
\caption {\label{fig3} Comparison between the key properties of the conventional (a)--(d) BDG, (e)--(h) FCE MG1, and (i)--(l) FCE MG2. (a), (c), and (i) depict the dielectric functions with respect to $z$. (b), (f), and (j) show the FEM-simulated dispersion relations. Blue and red insets illustrate the spatial electric field ($E_{y}$) distributions of the band edge modes on the $y=0$ plane. The vertical dotted lines represent the mirror planes in the computational cells. (c), (g), and (k) show the calculated radiative $Q$ factors of the lower and upper bands. (d), (h), and (l) show the transmission spectra through the BDG, FCE MG1, and FCE MG2. In the FEM simulations, we used the structural parameters $\epsilon_{\rm{av}}=4.00$, $\Delta\epsilon=1.00$, $\epsilon_{s}=1.00$, $t=0.30~\Lambda$, and $\rho = 0.40$. }
\end{figure*}

To analyse the superposition of scattering processes, owing to higher Fourier harmonic components, we introduce the resonance-assisted grating equation:
\begin{equation}\label{modified-phase-matching}
k_0 \sin \theta_{r,vn} = k_0 \sin \theta_{t,vn} = \Delta k_z \pm (2K  - vnK),
\end{equation} 	 	
where the $+$ and $-$ signs represent the propagating waves $P(x,z)$ and $N(x,z)$, respectively, $nK$ is the magnitude of the grating vector by the $n$th Fourier harmonic, and $v$ represents the order of the scattering processes. Figure~\ref{fig2} illustrates the resonant diffraction according to Eq.~(\ref{modified-phase-matching}) near the fourth stop band. Comparing Eq.~(\ref{modified-phase-matching}) and Eq.~(\ref{phase-matching}), it can be observed that Eq.~(\ref{phase-matching}) with $m=0$ is equivalent to Eq.~(\ref{modified-phase-matching}) with $(v, n)=(1,2)$ and $(2,1)$ for the $P$ and $N$ waves, respectively. For a higher diffraction order with $m=+1$, Eq.~(\ref{phase-matching}) is equivalent to Eq.~(\ref{modified-phase-matching}) with $(v, n)=(1,1)$ for the $P$ wave and $(v, n)=(1,3)$ and $(3,1)$ for the $N$ wave. Similarly, Eq.~(\ref{phase-matching}) with $m=-1$ is equivalent to Eq.~(\ref{modified-phase-matching}) with $(v, n)=(1,3)$ and $(3,1)$ for the $P$ wave and $(v, n)=(1,1)$ for the $N$ wave. As a result, we conclude that in the vicinity of the fourth stop band beyond the subwavelength limit, zero-order resonant diffraction is determined by the superposition of scattering processes, owing to the first and second Fourier harmonics, whereas higher diffraction orders with $m=\pm 1$ are determined by the first and third Fourier harmonics. Inspired by the analysis of resonance-assisted diffraction using Eq.~(\ref{modified-phase-matching}), we introduce and analyse Fourier-component-engineered (FCE) metagratings, namely FCE MG1 and FCE MG2, via rigorous finite element method (FEM) simulations. FCE MG1 does not possess the first and third Fourier harmonic components in the spatial dielectric function, and it is expected to realize highly efficient zero-order spectral responses beyond the subwavelength limit. By employing FCE MG2, which does not possess the first, second, and third Fourier harmonic components, we expect to obtain high-$Q$ bound states near the fourth stop band \cite{SGLee2021-1}.

In Fig.~\ref{fig3}, we compare the key properties of the conventional BDG and those of the corresponding FCE MG1 and FCE MG2. As shown in Figs.~\ref{fig3}(a), \ref{fig3}(e), and \ref{fig3}(i), while the conventional BDG has simple steplike dielectric functions $\epsilon_{\rm{BDG}} = \epsilon_a$ and $\epsilon_b$ when $|z| < \rho\Lambda /2$ and $|z| \geq \rho\Lambda /2$, respectively, the FCE MG1 and FCE MG2 has complex dielectric functions $\epsilon_{\rm{MG1}} = \epsilon_{\rm{BDG}} - \epsilon_{1} \cos (Kz) - \epsilon_{3} \cos (3Kz)$, and  $\epsilon_{\rm{MG2}} = \epsilon_{\rm{BDG}} - \epsilon_{1} \cos (Kz) - \epsilon_{2} \cos (2Kz) -\epsilon_{3} \cos (3Kz)$, respectively. The simulated dispersion relations shown in Figs.~\ref{fig3}(b), \ref{fig3}(f), and \ref{fig3}(j) show that the fourth bandgaps open at $k_z=0$ beyond the subwavelength limit ($\Omega_{\rm{Re}} > 1$) for the conventional BDG and FCE metagratings. The dispersion curves show the real parts of the eigenfrequencies in the Brillouin zone; those for the BDG and FCE metagratings seem similar. However, noticeable differences between the BDG, MG1, and MG2 can be observed from the spatial electric field ($E_y$) distributions in the insets. In the conventional BDG, both the lower and upper band edge modes with asymmetric (A) and symmetric (S) spatial electric field distributions are radiative out of the grating. At the second stop band in the subwavelength region, asymmetric edge modes become symmetry-protected BICs \cite{SGLee2019-1,Kazarinov1985}. However, at the fourth stop band beyond the subwavelength limit, even the asymmetric edge mode becomes radiative, owing to the higher diffraction orders with $m=\pm 1$, as shown in Fig.~\ref{fig3}(b). We note that protection by the symmetry mismatch is valid only for the zero-order diffraction radiating in the vertical direction. In FCE MG1 (shown in Fig.~\ref{fig3}(f)), while the symmetric upper band edge mode is radiative out of the grating, the asymmetric lower edge mode becomes the symmetry-protected BIC. This is because there are no higher diffraction orders, owing to the first and third Fourier harmonics. In FCE MG2 (shown in Fig.~\ref{fig3}(j)), the symmetric upper band and asymmetric lower band edge modes are strongly localized in the grating layer because there is no radiative scattering by the higher Fourier harmonics. The effects of the higher Fourier harmonic components are similarly observed by investigating the radiative $Q$ factors, which are plotted in Figs.~\ref{fig3}(c), \ref{fig3}(g), and \ref{fig3}(k). In the BDG, Bloch modes in both the lower and upper band branches have low $Q$ values ($\sim 10^3$), and no BIC is observed. In FCE MG1, the BIC in the lower band exhibits a $Q$ factor that is larger than $10^8$ at the $\Gamma$ point; however, the $Q$ values decrease rapidly as $k_z$ moves away from the $\Gamma$ point. In FCE MG2 (shown in Fig.~\ref{fig3}(k)), the Bloch modes in both the lower and upper band branches become strongly confined BICs with high $Q$ values ($\sim 10^8$) in the computational range of $|k_z|\leq 0.07~K$. The $Q$ factors in FCE MG2 are approximately $10^5$ times larger than those in the conventional BDG with the same lattice parameters, except for the profile of the dielectric function.

Figures~\ref{fig3}(d), \ref{fig3}(h), and \ref{fig3}(l) illustrate the transmission spectra through the conventional BDG, FCE MG1, and FCE MG2, respectively, for two different incident angles $\theta =0^\circ$ and $1^\circ$. The grating structures exhibit three transmitted diffraction orders: $m=-1$, $0$, and $+1$. The transmittance curves for $m=-1$ and $+1$, shown in blue and green lines, respectively, are the same when $\theta =0^\circ$; however, they can be distinguished when $\theta =1^\circ$. As illustrated in Fig.~\ref{fig3}(d), no diffraction with $100\%$ efficiency is observed for the conventional BDG. However, as illustrated in Fig.~\ref{fig3}(h), zero-order transmittance curves (red lines) through FCE MG1 exhibit the ordinary profile of Fano resonances \cite{AEMiroshnichenko2010}, in which transmitted diffraction efficiency rapidly varies from $0\%$ to $100\%$ in the vicinity of resonant frequencies, irrespective of the incident angle $\theta$. At normal incidence with $\theta =0^\circ$, the zero-order transmittance curve exhibits only the low-$Q$ resonance by the upper band edge mode because the embedded BIC in the lower band edge mode (shown by a blue solid circle in the transmittance curve) does not generate the resonance effect. When $\theta =1^\circ$, the simulated transmittance curve for $m=0$ exhibits a low- and high-$Q$ resonance by the upper and lower band mode, respectively. Diffraction orders with $m=-1$ and $+1$ exhibit only zero transmittance through the FCE MG1. As illustrated in Fig.~\ref{fig3}(l), when $\theta =0^\circ$, the zero-order transmittance curve through FCE MG2 exhibits a high-$Q$ resonance, in which diffraction efficiency varies from $0\%$ to $97\%$, and the symmetry-protected BIC is not associated with resonance. When $\theta =1^\circ$, two quasi-BICs are observed in the transmittance curve for $m=0$. In the vicinity of the quasi-BIC by the lower and upper band mode, the zero-order transmission efficiency varies from $18\%$ to $79\%$ and from $0\%$ to $96\%$, respectively.

\begin{figure}[]
\centering \includegraphics[width=8.0cm]{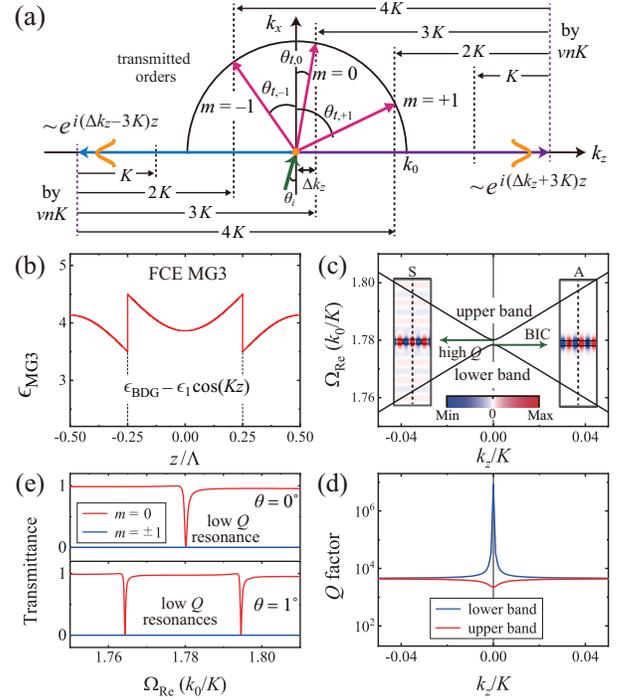}
\caption {\label{fig4} Resonant diffraction near the sixth stop band beyond the subwavelength limit. (a) Diffraction orders by Eq.~(\ref{modified-phase-matching-2}). (b) Spatial dielectric function with respect to $z$. Simulated (c) dispersion relations, (d) radiative $Q$ factors, and (e) transmission spectra near the sixth stop band of FCE MG3. In the simulations, we used the grating parameters $\epsilon_{\rm{av}}=4.00$, $\Delta\epsilon=1.00$, $\epsilon_{s}=1.00$, $t=0.15~\Lambda$, and $\rho = 0.50$. }
\end{figure}

Analysis using Eq.~(\ref{modified-phase-matching}) is valid only in the vicinity of the fourth stop bands open at the third-order $\Gamma$ point. We now generalize the superposition of scattering processes by investigating the resonant diffraction near the sixth stop band open at the fourth-order $\Gamma$ point ($k_z=3~K$ in the extended Brillouin zone). With the electric field distribution $E_y(x,z)=P(x,y) + N(x,y) + E_{\rm{rad}}$, where $P(x,y)= A\exp[i( \Delta k_z+ 3K )z]\varphi(x)$ and $N(x,z)=B\exp[i( \Delta k_z- 3K )z]\varphi(x)$, the resonance-assisted grating equation can be written as
\begin{equation}\label{modified-phase-matching-2}
k_0 \sin \theta_{r,vn} = k_0 \sin \theta_{t,vn} = \Delta k_z \pm (3K  - vnK).
\end{equation} 	 	
Figure~\ref{fig4}(a) illustrates the resonant diffraction using Eq.~(\ref{modified-phase-matching-2}). We note that Eq.~(\ref{phase-matching}) with $m=0$ is equivalent to Eq.~(\ref{modified-phase-matching-2}) with $(v, n)=(1,3)$ and $(3,1)$ for both the $P$ and $N$ waves, respectively. For a diffraction order of $m=+1$, Eq.~(\ref{phase-matching}) is equivalent to Eq.~(\ref{modified-phase-matching-2}) with $(v, n)=(1,2)$ and $(2,1)$ for the $P$ wave and $(v, n)=(1,4)$, $(4,1)$, and $(2,2)$ for the $N$ wave. Likewise, Eq.~(\ref{phase-matching}) with $m=-1$ is equivalent to Eq.~(\ref{modified-phase-matching-2}) with $(v, n)=(1,4)$, $(4,1)$, and $(2,2)$ for the $P$ wave and $(v, n)=(1,2)$ and $(2,1)$ for the $N$ wave. To summarize, the zero-order resonant diffraction, shown in Fig.~\ref{fig4}(a), is determined by the superposition of scattering processes, owing to the first and third Fourier harmonics, whereas higher diffraction orders with $m=\pm 1$ are determined by the first, second, and fourth Fourier harmonics.

\begin{figure}[b]
\centering \includegraphics[width=8.5 cm]{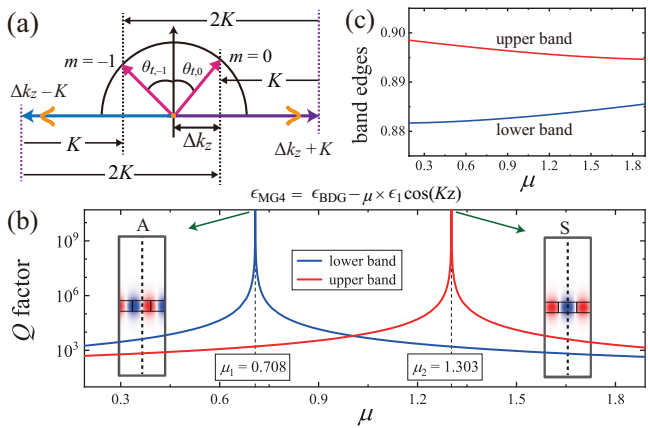}
\caption {\label{fig5} Resonant diffraction in the vicinity of the third stop band beyond the subwavelength limit. (a) Diffraction orders by Eq.~(\ref{modified-phase-matching-3}). Simulated (b) radiative $Q$ factors and (c) eigenfrequencies of the two band-edge modes as a function $\mu$. In addition to the engineered dielectric function, the structural parameters were the same as those in Fig.~\ref{fig2}.}
\end{figure}

To verify the resonant diffraction using Eq.~(\ref{modified-phase-matching-2}), we consider an additional metagrating, FCE MG3, which has the engineered dielectric function $\epsilon_{\rm{MG3}} = \epsilon_{\rm{BDG}} - \epsilon_{1} \cos (Kz) $, as shown in Fig.~\ref{fig4}(b)]. Analysis using Eq.~(\ref{modified-phase-matching-2}) indicates that higher diffraction orders of $m=\pm 1$ can be suppressed by eliminating the first, second, and fourth Fourier components from $\epsilon_{\rm{BDG}}$. To avoid complex spatial profiles of the dielectric constant, we set the second and fourth Fourier coefficients, $\epsilon_{2}$ and $\epsilon_{4}$, to zero using the lattice parameter $\rho = 0.5$, instead of eliminating the spatial modulations $\epsilon_{2} \cos (2Kz)$ and $\epsilon_{4} \cos (4Kz)$. The simulated dispersion curves and radiative $Q$ factors are shown in Fig.~\ref{fig4}(c) and Fig.~\ref{fig4}(d), respectively; they indicate that the fourth band gap opens at the fourth-order $\Gamma$ point. The lower band edge mode with an asymmetric field distribution becomes a high-$Q$ BIC, whereas the upper band edge mode with a symmetric field distribution radiates out of the grating layer. The spatial electric field distribution of the symmetric edge mode is illustrated in the inset of Fig.~\ref{fig4}(c); it indicates that out-of-plane radiation occurs only in the zero-order vertical direction because the higher diffraction orders are suppressed through the engineering of Fourier harmonic components. The simulated transmittance curves for FCE MG3 are shown in Fig.~\ref{fig4}(e). These curves demonstrate that irrespective of the incident angle $\theta$, diffraction orders of $m=\pm 1$ exhibit only zero transmittance, whereas the zero-order spectral responses vary from 0 to 1 in the vicinity of the resonance frequencies beyond the subwavelength limit.

Next, we consider the diffraction in the vicinity of the third stop band in the grey region, where two diffraction orders of $m=0$ and $m=-1$ coexist. Using the resonance-assisted grating equation, given by
\begin{equation}\label{modified-phase-matching-3}
k_0 \sin \theta_{r,vn} = k_0 \sin \theta_{t,vn} = \Delta k_z \pm (K  - vnK),
\end{equation} 	 	
and the corresponding phase-matching processes illustrated in Fig.~\ref{fig5}(a), we conclude the following: in the vicinity of the third stop band, both diffraction orders, $m=0$ and $m=-1$, are determined by the superposition of scattering processes, owing to the first and second Fourier harmonic components. Thus, the diffraction order of $m = -1$ cannot be selectively suppressed through the engineering of the Fourier harmonic components. However, high-$Q$ BICs can be achieved by utilizing the engineered dielectric function $\epsilon_{\rm{MG4}}= \epsilon_{\rm{BDG}} - \epsilon_{1} \cos (Kz)- \epsilon_{2} \cos (2Kz)$ or $\epsilon_{\rm{BDG}} - \epsilon_{1} \cos (Kz)$ with $\rho = 0.5$. As an additional example of the engineering of Fourier harmonic components, we introduce a metagrating, FCE MG4, that exhibits a high-$Q$ BIC at one edge of the third stop band. The dielectric function of FCE MG4 was set to $\epsilon_{\rm{MG4}}= \epsilon_{\rm{BDG}} - \mu\times \epsilon_{1} \cos (Kz)$, where the coefficient $\mu$ is introduced to control the strength of the first Fourier harmonic component. By varying the value of $\mu$, as shown in Fig.~\ref{fig5}(b), there exist two critical values $\mu_1 = 0.708$ and $\mu_2 = 1.303$, where out-of-plane radiations, owing to the first and second Fourier harmonic components, eliminate each other via destructive interference. When $\mu = \mu_1$ ($\mu = \mu_2$), the BIC with an asymmetric (symmetric) field distribution is observed in the lower (upper) band edge. There are two critical values, $\mu_1 < 1$ and $\mu_2 > 1$, because the contribution of the first Fourier component increases as the value of $\mu$ increases or decreases from 1. We consider that the BIC is observed at one of the band edges because the condition for complete destructive interference strongly depends on the eigenfrequency of the band edge mode. Figure~\ref{fig5}(c) shows that the band edge frequencies vary as $\mu$ increases.

\begin{figure}[]
\centering \includegraphics[width=8.5 cm]{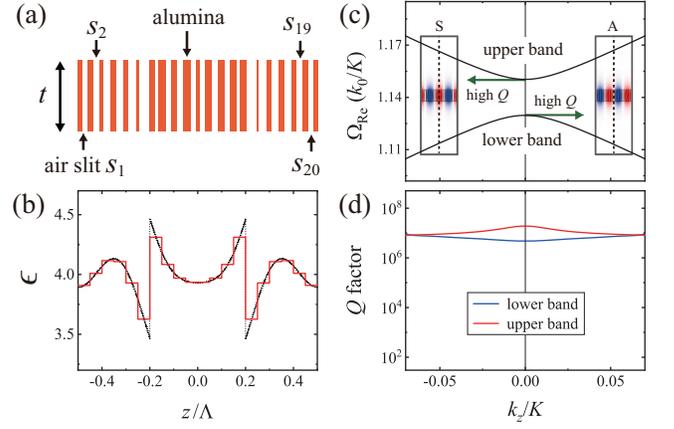}
\caption {\label{fig6} Metagrating utilizing the metamaterial concept. (a) Schematic of an alumina-air metasurface with $t=0.30~\Lambda$. (b) Effective dielectric constant of the alumina-air metasurface in the computational unit cell of size $\Lambda$. The dotted line represents the dielectric constant of the FCE MG2. Simulated dispersion relations (c) and radiative $Q$ factors (d) near the fourth stop band of the alumina-air metasurface. }
\end{figure}

In this study, we performed analytical and numerical investigations to control light diffraction beyond the subwavelength limit. The proposed FCE metagratings may be practically implemented by utilizing the metamaterial concept \cite{Schurig2006,HFMa2010}, which can mimic the required dielectric constant profile using a series of discrete step functions \cite{SGLee2021-1,Haggans1993}. Owing to the complex spatial variations, it is challenging to implement FCE metagratings at optical wavelengths. However, current microfabrication technology is sufficient to demonstrate the effects of engineered dielectric functions in the microwave range. To visualize this, as shown in Fig.~\ref{fig6}(a), we consider an alumina-air metasurface consisting of air slits, $S_{j}$, in an alumina slab ($\epsilon_{\mathrm{al}}=9.7$ at 10 GHz). Twenty air slits are appropriately located in the unit cell of size $\Lambda$, such that the effective dielectric function of the alulmina-air metasurface mimics that of the FCE MG2, as illustrated in Fig.~\ref{fig6}(b). Even though the effective dielectric function of the alumina-air metasurface varies in discrete steps, not only the dispersion curves of the alumina-air metasurface in Fig.~\ref{fig6}(c) are nearly the same as those of the FCE MG2 in Fig.~\ref{fig3}(j), but also the radiative $Q$ factors in the alumina-air metasurface in Fig.~\ref{fig6}(d) are comparable to the high $Q$ values in the FCE MG2 in Fig.~\ref{fig3}(k). The radiative $Q$ factors in the alumina-air metasurface are approximately $10^4$ times larger than those in the conventional BDG in Fig.~\ref{fig3}(c).

In conclusion, we introduced resonance-assisted grating equations and analysed the resonant diffraction phenomena beyond the subwavelength limit, where multiple diffraction orders coexist. Our analysis shows that the resonant diffraction is governed by the superposition of scattering processes, owing to the higher Fourier harmonic components in the grating parameters. By appropriately engineering the higher Fourier harmonic components, unwanted diffraction orders can be suppressed, and important resonance-assisted physical phenomena, such as BICs and zero-order spectral responses with $100\%$ diffraction efficiency, can be achieved, even beyond the subwavelength limit. Four types of metagratings were introduced as examples of the engineering of Fourier harmonic components. They were analysed in the vicinity of the third, fourth, and sixth stop bands. The proposed metagrating concepts have the capability to control unwanted diffraction orders; thus, they may be applied to operations beyond the subwavelength limit.

\newpage

\section*{Methods}
%\textbf{Methods} \\ 
\noindent To verify the resonant diffraction governed by the superposition of scattering processes owing to higher Fourier harmonic components, one-dimensional (1D) lattices were investigated through the rigorous finite element method (FEM) simulations. We used the commercial software COMSOL multiphysics 5.3a. Since the 1D periodic structures are invariant in the $y$-direction, simulations were performed in 2D $xz$-plane. Computational cell of size $\Lambda \times 15~\Lambda$ was employed to obtain dispersion relations, $Q$ factors, spatial electric field distributions, and transmission spectra. Bloch periodic boundary condition was used in the $x$-direction and perfectly matched layer absorbing boundary condition was employed in the $z$-direction. For reliable simulations, we used user-controlled mesh with extremely fine element size.

\section*{Data availability}
%\textbf{Data availability} \\ 
\noindent Source data are provided with this paper. All other data that support the plots within this paper and other findings of this study are available from the corresponding author upon reasonable request.
\section*{acknowledgments}
%\textbf{acknowledgments} \\ 
\noindent This research was supported by grants from the National Research Foundation of Korea funded by the Ministry of Education (No. 2020R1I1A1A01073945) and Ministry of Science and ICT (No. 2020R1F1A1050227), along with the Gwangju Institute of Science and Technology Research Institute.
\section*{Author contributions}
%\textbf{Author contributions} \\ 
\noindent The resonance-assisted grating equations were developed by S.-G, and numerical simulations were performed by S.-G. and S.-H. C.-K. analysed the numerical data. All authors contributed to discussions and manuscript writing.
\section*{competing interests}
%\textbf{competing interests} \\ 
\noindent The authors declare no competing interests.
\section*{Additional information}
%\textbf{Additional information} \\ 
\noindent Correspondence and requests for materials should be addressed to S.-G. and C.-K.

\end{document}